\documentclass{llncs}
\usepackage{graphicx}	
\usepackage{placeins}
\usepackage{amsfonts}		
\usepackage{amssymb}
\usepackage{siunitx,chemmacros}		
\usepackage{comment}
\usepackage{seqsplit}
\usepackage{booktabs}		
\usepackage{multirow}		
\usepackage{tabularx}		
\usepackage{tabulary}		
\usepackage{longtable}		
\usepackage{ltxtable}		
\usepackage{caption}

\newcommand{\gene}[1]{\textit{#1}}		
\newcommand{\specie}[1]{\textit{#1}}		

\begin{document}

\frontmatter          
\pagestyle{headings}  
\mainmatter              
\title{Experimental Analysis of XPCR-based protocols}
\titlerunning{XPCR Exp Analysis}  
\author{Giuditta Franco \and Francesco Bellamoli \and Silvia Lampis}
\authorrunning{Francesco Bellamoli et al.} 
\tocauthor{Giuditta Franco, Francesco Bellamoli, Silvia Lampis}

\institute{Universit\`{a} degli Studi di Verona, Italy,\\
\email{giuditta.franco@univr.it, francesco.bellamoli@gmail.com, silvia.lampis@univr.it}
}

\maketitle              

\begin{abstract}

This paper reports some experimental results validating in a broader context a variant of PCR, called  XPCR, previously introduced and tested on relatively short synthetic DNA sequences. Basic XPCR technique confirmed to work as expected, to concatenate two genes of different lengths, while a library of all permutations of three different genes (extracted from the bacterial strain {\it Bulkolderia fungorum} DBT1) has been realized in one step by multiple XPCR. 

Limits and potentialities of the protocols have been discussed, and tested in several experimental conditions, by aside showing that overlap concatenation of multiple copies of one only gene is not realizable by these procedures, due to strand displacement phenomena. In this case, in fact, one copy of the gene is obtained as a unique amplification product.

\keywords{Models for biotechnological protocols, multiple XPCR method, PCR amplification, polymerase extension, unexpected outcomes.}
\end{abstract}

\section{Introduction}

Polymerase Chain Reaction (PCR) is a versatile DNA amplification technology introduced by Karry Mullis in the 1980s, that revolutionized molecular biology. Since then PCR applications have reached diverse fields, including ecology,
criminal forensics, food science, and diagnostic medicine~\cite{Bartlett:03}. From an algorithmical perspective, the technique shows interesting combinatorial properties~\cite{Manca:08}. However, when used in non-standard ways, it still yields unexpected and very complex behaviors~\cite{Holland:91,Kanagawa:03,Pavlov:04}, anomalies often ascribed to ``experimental noise''. 

As a matter of fact, several standard biomolecular protocols do not ensure the needed precision to compute with DNA. An intriguing task in molecular computing is to characterize new procedures and protocols, within the scope of given experimental limitations, in order to be sufficiently reliable, controllable and predictable, for both molecular computation and biological applications. Several new or improved methodologies have been proposed in the context of DNA recombinant technology, namely including Y operation~\cite{shapiro} and XPCR technique~\cite{Franco:11}, which selectively concatenate DNA sequences. These biotechnological procedures are still going through several adjustments, in order to improve their efficiency, to optimize their performance, their reproducibility, and to delimit their implementation potentialities. 

This paper reports some experiments which broaden the application field of XPCR technique (in both its basic and multiple variants~\cite{Franco:11}) to genetic sequences. In other terms, after a previous validation on relatively short synthetic sequences (150 bp long), the technique is here experimentally validated on three genes (natural sequences long 311, 518, and 1019 bp) extracted from an environmental bacterial strain, the {\it Bulkolderia fungorum} DBT1~\cite{DiGregorio:04}, having remarkable PAH (polycyclic aromatic hydrocarbon) degradation capabilities. Moreover, some open questions about experimental implementation of XPCR technique, which were posed in~\cite{Franco:05,Franco:11}, have been here answered. Importantly enough, an unexpected outcome was amplified in the experiments designed to concatenate two copies of one same gene, as reported in section~\ref{sub2}. XPCR amplification of fragments $\alpha A \gamma$ and $\gamma A \beta$ produced $\alpha A \gamma$, in all experimental settings (including the use of Pfu polymerase), rather than the expected $\alpha A \gamma A \beta$ (the overlap product of the two input molecules).

In this work, concatenation of two different genes and a library with all permutations of the three genes above have been successfully generated by XPCR-based protocols. In next section XPCR protocol is recalled, as it is known at the state of the art, while in section~\ref{sec:2} and section~\ref{sec:3} new results are described. In particular, in section~\ref{sec:2} the concatenation between couples of different genes have been realized, while in subsection~\ref{sub2} the impossibility to concatenate two copies of a same gene by XPCR is shown and discussed. In section~\ref{sec:3} a ``multiple'' concatenation of three (different) genes has been produced, while in subsection~\ref{sub3} the necessity to have different overlapping primers between couples of genes is outlined and discussed. Final sections conclude the paper, along with a description of experimental materials and methods.

As for the notation, we denote the genes with Capital letters $A$, $B$, $D$, and primers with Greek letters, such as $\alpha, \beta, \gamma$ and $\delta$, under the assumption that concatenation of sequences is indicated by juxtapposition of letters. In particular, $\alpha X \beta$, with $X$ being one of the genes, represent specific sequences where gene X has been elongated by a prefix $\alpha$ and a suffix $\beta$. Finally, overlined letters represent reversed complementary sequences, such as, for example, the backward primer $\overline{\beta}$, and arrow tips in figures denote 3$^\prime$ extremities of molecules.

\section{XPCR protocol}
Recombination of two DNA strings along with a (given) common substring $\gamma$ may be efficiently implemented by XPCR$_\gamma$ (or $\gamma$-cross-pairing PCR~\cite{Franco:05b}), as well as other procedures (such as DNA extraction, combinatorial libraries generation, and mutagenesis~\cite{Franco:05b,Franco:05,Franco:06,Manca:08,Franco:11}) which can be described in terms of {\it null context splicing rules}. These are rewriting rules introduced by T. Head in the context of Formal Language Theory (FLT) as particular H-systems, inspired by selective recombination induced by few (existent in nature) restriction enzymes~\cite{head}: 
$$ \alpha X \gamma Y \beta, \alpha V \gamma W \beta \rightarrow \alpha X \gamma W \beta, \alpha V \gamma Y \beta $$
Such rule may be seen as a specific chimeras generation method, which we briefly call $\gamma$-recombination. It cleary affects only strings containing a given sequence $\gamma$, that are technically (in FLT) called $\gamma$-superstrings. 

\begin{figure}[h] 
	\centering
	\includegraphics[width=.77\columnwidth]{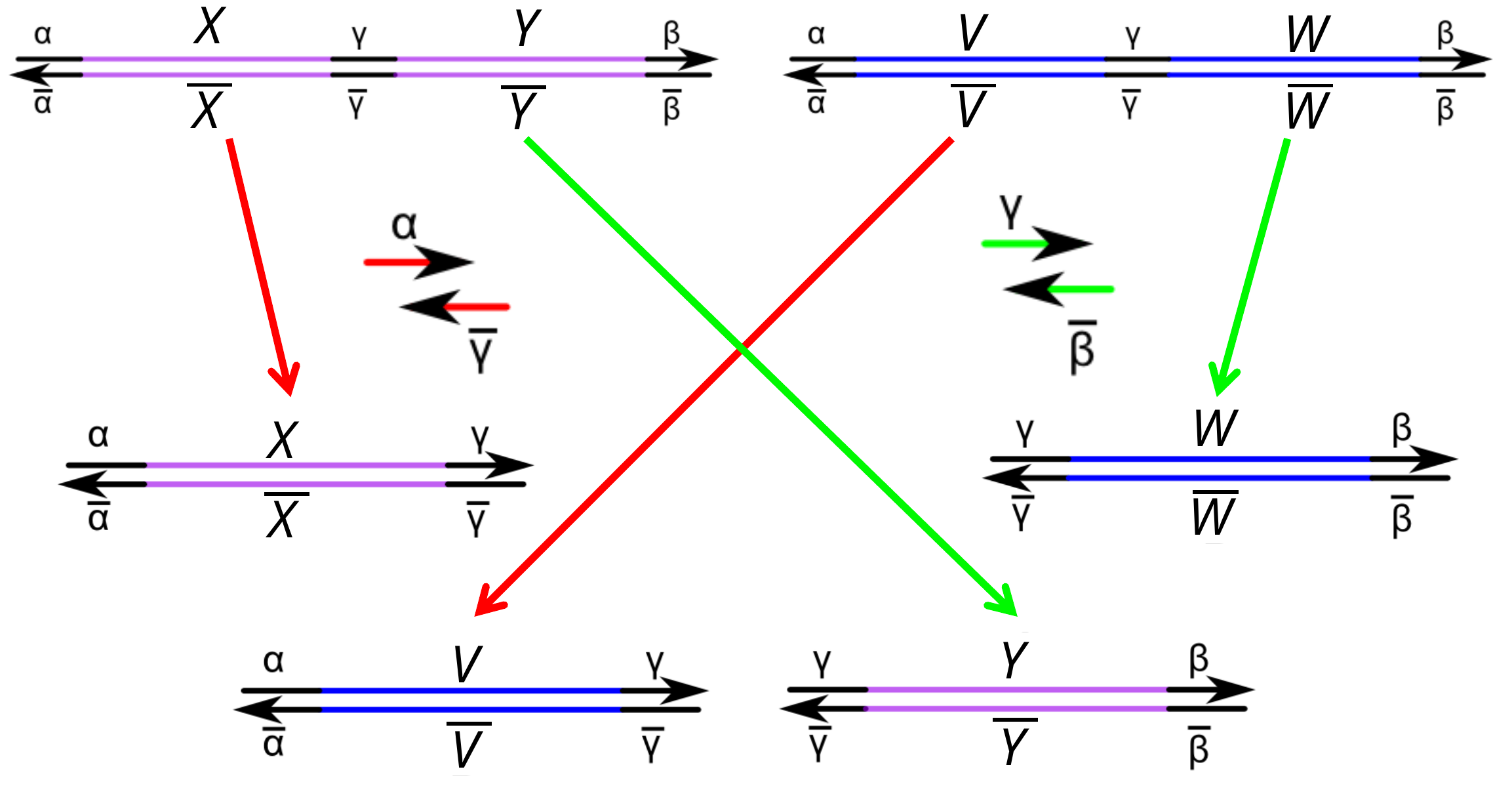}
	\caption{Preparatory step for XPCR$_\gamma$.}
	\label{fig:xpcrcut}
\end{figure}

\begin{figure}[ht] 
	\centering
	\includegraphics[width=.85\columnwidth]{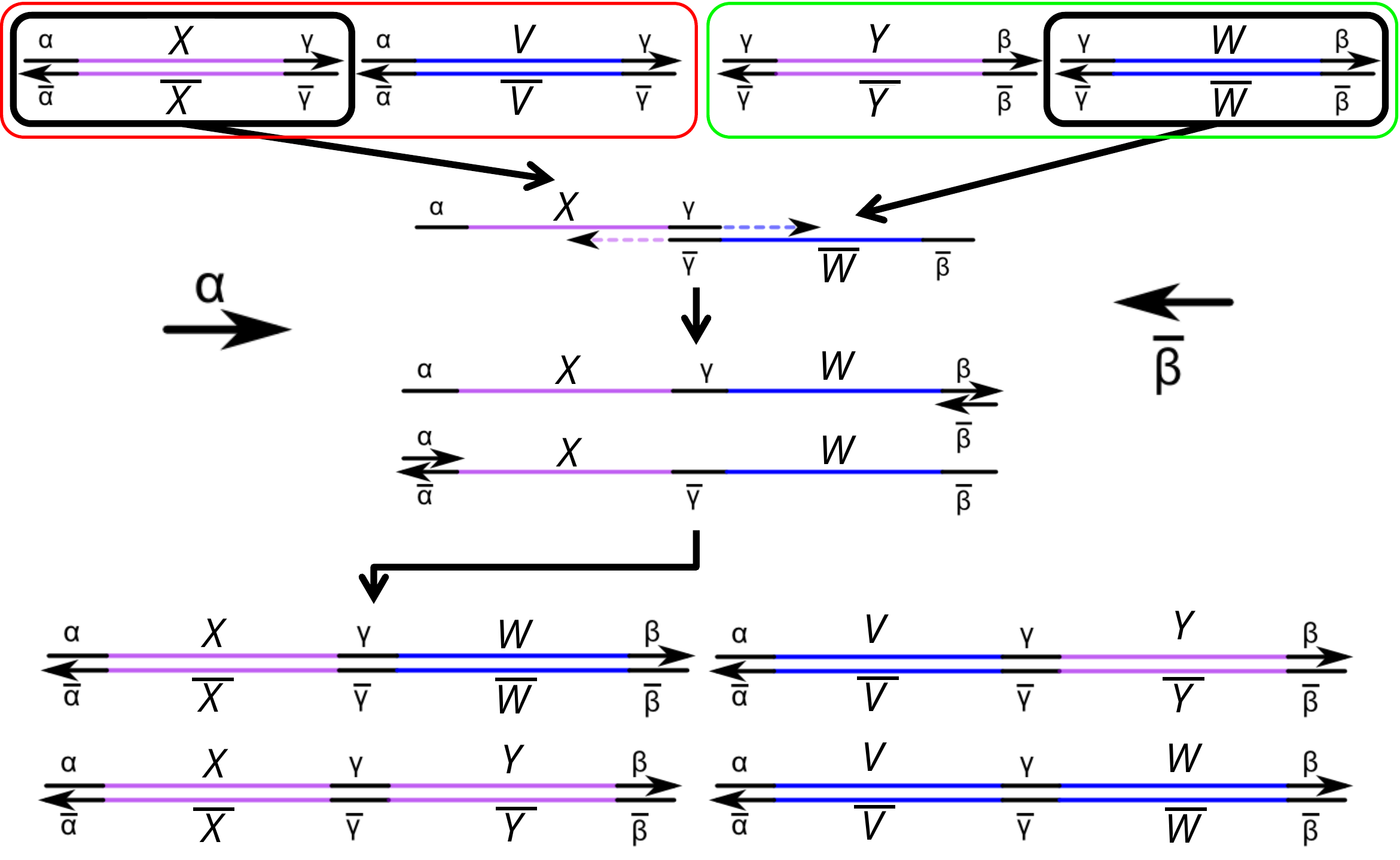}
	\caption{Recombination step of XPCR$_\gamma$.}
	\label{fig:xpcrbasic}
\end{figure}

After the rule application, the left and right side contexts of $\gamma$, in all the $\gamma$-superstrings in the pool, are recombined in all possible ways by keeping the (left, right) order. See Figure~\ref{fig:xpcrcut}) and Figure~\ref{fig:xpcrbasic}), where the composition of sequences $X, Y, V, W$ preceding and following $\gamma$ is usuallyn unknown. 

Once $\alpha$, $\beta$ and $\gamma$ sequences are designed (or given), XPCR$_\gamma$ procedure may be applied to any
heterogeneous DNA pool composed of $\alpha$-prefixed and $\beta$-suffixed double-stranded sequences, to produce all possible $\gamma$-chimeras as additional molecules. The technique was successfully tested in vitro, under several experimental conditions, also as an implementation basis of several other algorithms of interest in biotechnology and diagnostics~\cite{Franco:11}. However, experiments from the literature were all set on synthetic molecules, and the lengths tested were 15 bp for primers and recombinant portions, and few hundreds for longest chimeras. A main point in this paper is to test the technique for very long (genetic) sequences.

In the following a short description of the lab protocol is reported. The initial pool P of DNA molecules is split in two pools, P$_1$ and P$_2$, where respectively, a PCR si performed, with primers ($\alpha$, $\overline{\gamma}$) and ($\gamma$, $\overline{\beta}$). At this poin an electrophoresis allows us to remove unwanted products (initial molecules), and to isolate the desired (shorter than previously present) $\gamma$-suffixed and
$\gamma$-prefixed sequences just amplified by PCR. This step is referred to as a preparatory step for XPCR$_\gamma$, depicted in Figure~\ref{fig:xpcrcut}. It provide us with strings containing the sequences on the left and on the right side of $\gamma$, from all molecules.

Purified products are mixed together into a new pool P, where a regular PCR is performed, with ($\alpha$, $\overline{\beta}$)
primers, by realizing the recombinant process depicted in Figure~\ref{fig:xpcrbasic}, where each
primer anneals to a different template and it is extended, while $\gamma$ and
$\overline{\gamma}$ sequences located on different templates annealing to each other and being extended by DNA
polymerase. This step forms recombined sequences, which are $\alpha$-prefixed and $\beta$-suffixed chimeras containing $\gamma$. 

\begin{figure}[h!] 
	\centering
	\includegraphics[width=.78\columnwidth]{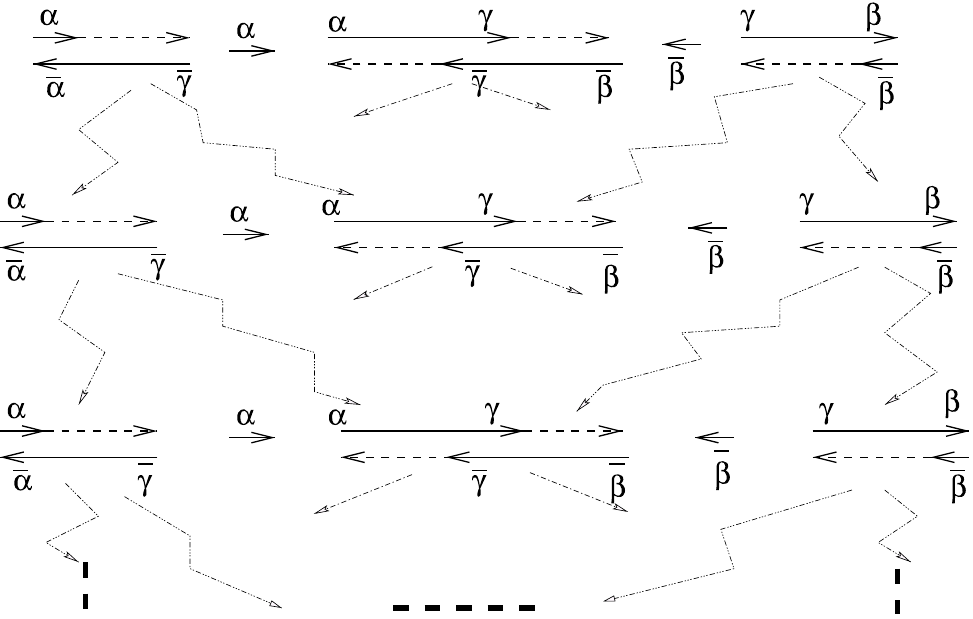}
	\caption{Amplification by means of XPCR$_\gamma$ (from \cite{Franco:06}). Long strands in the middle are seeds of exponential
	amplification, newly produced at every iteration within PCR.}
	\label{fig:xpcramp}
\end{figure}

Seeds for exponential amplification with primers $\alpha$ and
$\overline{\beta}$ are generated at every iteration of the PCR implementing the recombination step in Figure~\ref{fig:xpcrbasic}, hence XPCR$_\gamma$ results in a superexponential generation of $\gamma$-chimeras (very efficient method, with no production of redundant material) as sketched in Figure~\ref{fig:xpcramp}. 

Multiple XPCR or $n$-XPCR was designed~\cite{Franco:05} and tested in~\cite{Franco:11} as a variant of XPCR, to realize a parallel concatenation of $n$ different types of
double-stranded DNA sequences in a single step. Beside other experimental validations, starting from a pool P with seven fragments $\{\alpha\delta_1 \gamma_1, \gamma_i\delta_{i+1} \gamma_{i+1}, \gamma_7\delta_7\beta\  \mid i=1,..,6\}$, molecule $\alpha\delta_1\gamma_1\delta_2 \gamma_2 \delta_3 \gamma_3 \delta_4 \gamma_4 \delta_5\gamma_5 \delta_6 \gamma_6 \delta_7 \beta$ (long 538 bp) with seven different binary forms $\delta$ and seven different overalapping primes $\gamma$ has been obtained in~\cite{Franco:11}.

In the next sections we will show some new experiments, proving that: $i)$ XPCR works well even on long natural sequences, to form chimeras of two different genes (Section~\ref{sec:2}), and $ii)$ multiple XPCR works well to recombine three different genes, resulting in the one-step-generation of a library with sequences about 2 kbp long (Section~\ref{sec:3}). Such experimentally validated extensions of XPCR broaden the relevance and the field of application of the technique. However, interestingly enough, in this same work new and clear limits of the protocol arose repetitively from the experiments: $i)$ XPCR does not implement the overlap recombination to concatenate two copies of the same (genetic) fragment, actually starting from input molecules $\alpha A \gamma$ and $\gamma A \beta$ it performs a sort of ``molecular fusion'', by amplifying the product  $\alpha A \gamma$ (subsection~\ref{sub2}), and $ii)$ multiple XPCR necessarly requires two different overalapping primers to recombine three genes, as with equal primers it does not work as expected (subsection~\ref{sub3}).

Bacterial strain \specie{Burkholderia fungorum} DBT1 \cite{DiGregorio:04} was used as
source for extracting the genes \gene{dbtAa} (1019 bp, ferredoxyn reductase), \gene{dbtAb} (311 bp, ferredoxyn),
and \gene{dbtAd} (518 bp, dioxygenase $\beta$ subunit), utilized in following experiments, and briefly called~\gene{A}, \gene{B}, and \gene{D}, respectively. Various sets of oligonucletides were designed and checked, including primers for extraction of genes and gene CDS, overhangs used to link sequences to genes, thus to create the ``building blocks'' of input pools, and primers $\alpha$, $\beta$, $\gamma$, $\delta$.

\section{XPCR-based two genes concatenation}\label{sec:2}
\begin{figure}[ht] 
	\centering
	\includegraphics[width=.59\columnwidth]{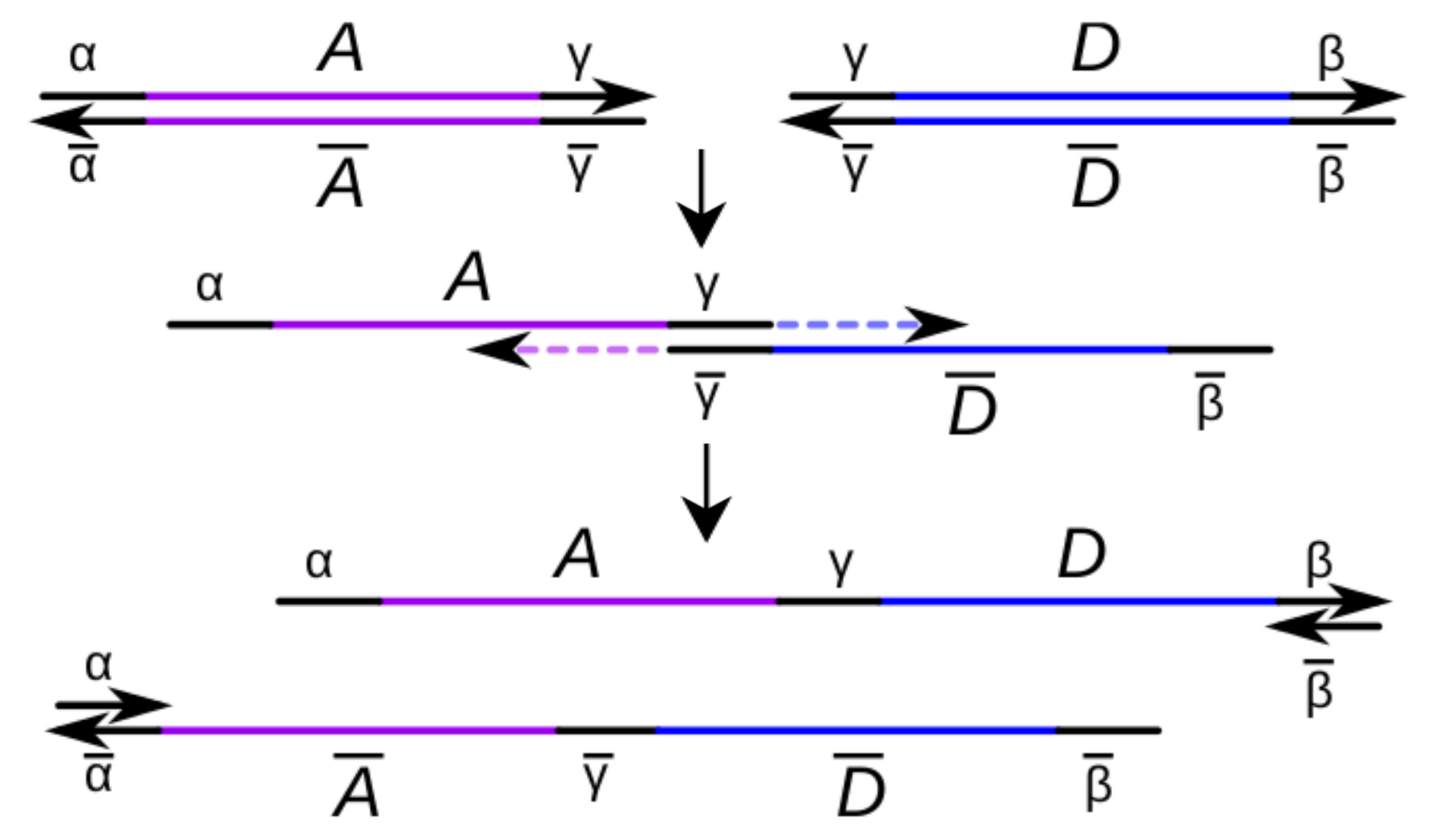}
	\caption{Concatenation of genes A and D by overlapping primer $\gamma$. Starting templates $\alpha A \gamma$ and $\gamma D
	\beta$, after denaturation, anneal along common $\gamma$ and are extended by DNA polymerase, leading to the concatenation $A\gamma D$, then amplified by primers  ($\alpha$, $\overline{\beta}$).}
	\label{fig:2gene_drw}
\end{figure}

Here we proved that the basic recombination step of $XPCR_\gamma$ (see Figure~\ref{fig:xpcrbasic}), simply depicted in Figure~\ref{fig:2gene_drw} over sequences $A$ and $D$, works correctly also for a couple of natural genes of relevant and different lengths. Our experimentl tests recombined any couple of the three genes $A, B, D$.

\begin{figure}[ht] 
	\centering
	\includegraphics[width=.99\columnwidth]{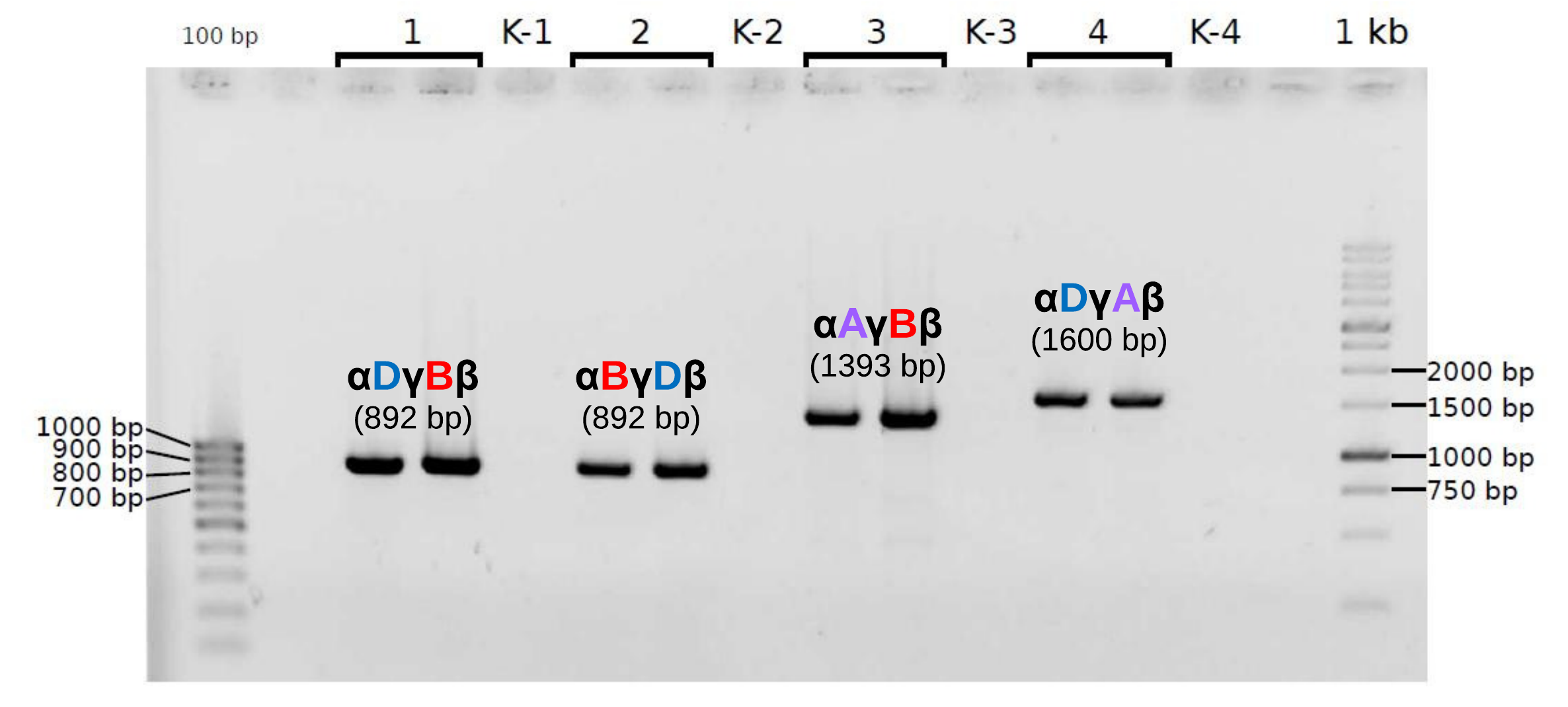}
	\caption{Successful two genes concatenation (by using \gene{Pfu} polymerase). Lanes 1, 2, 3, 4: templates $\alpha D \gamma$ + $\gamma B \beta$, $\alpha B \gamma$ +
	$\gamma D \beta$, $\alpha A \gamma$ + $\gamma B \beta$ and $\alpha D \gamma$ + $\gamma A \beta$, respectively. Lanes K-1, K-2, K-3, K-4: respective negative controls without any template.}
	\label{fig:2gene_3}
\end{figure}

Experiments of PCR reactions with primers ($\alpha$, $\overline{\beta}$) were performed to concatenate two out of three genes, $(A, B)$, $(D, A)$, and $(D, B)$, also in the reverse order $(B, D)$: expected results are reported in Figure~\ref{fig:2gene_3}, with products of the expected lengths: 892~bp for B and D concatenation, 1393~bp for gene \gene{A} concatenated to gene \gene{B}, and 1600~bp for the concatenation of D and A\footnote{Results were consistent with expected lengths, both using \gene{Taq} and \gene{Pfu} polymerase.}.

XPCR based protocol worked as hypothesized, to perform overlap concatenation of two genes, if and only if the two templates in each reaction contain different genes. Indeed, experiments which were attempted to concatenate two copies of one same gene $X$ by XPCR$_\gamma$, starting from  two sequences $\alpha X \gamma$ and $\gamma X \beta $, consistently produced $\alpha X \beta $ as only amplicon (details in the following subsection).

\subsection{Concatenation of two copies of one gene} \label{sub2}

\begin{figure}[h] 
	\centering
	\includegraphics[width=.89\columnwidth]{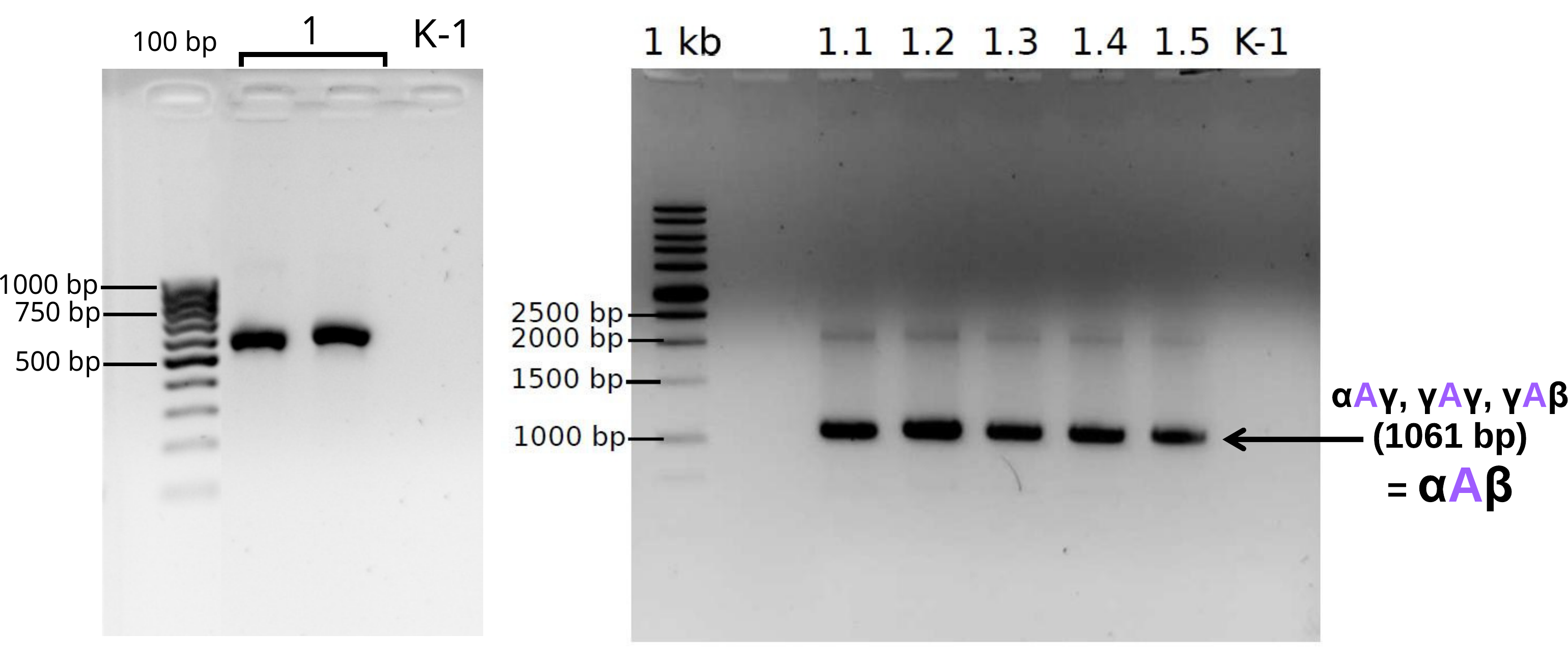}
	\caption{Attempt to concatenate two and three copies of one gene (by~\gene{Pfu} polymerase). Lane 1: templates $\alpha D \gamma$ + $\gamma D \beta$, exhibited product of about 500 bp. Lanes 1.1,
	1.2, 1.3, 1,4, 1.5: templates $\alpha A \gamma$ +
	$\gamma A \gamma$ + $\gamma A \beta$, with $\gamma A \gamma$ concentration ratio with respect to the others 10:1, and respectively different annealing
	temperatures. Lane K-1:
	negative control without any template.}
	\label{fig:3samegene_a_2}
\end{figure}

Experiments of PCR reactions with primers ($\alpha$, $\overline{\beta}$) over templates $\alpha X \gamma$ + $\gamma
X \beta$ and $\alpha X \gamma$ + $\gamma X \gamma$ + $\gamma X \beta$ were performed, to concatenate respectively two and three copies of one gene $X$ (out of the three), and the main (new, unexpected) product $\alpha X \beta $ has been obtained. A typical reaction product is shown in
Fig.\ref{fig:3samegene_a_2}. For example, on the right gel of Figure~\ref{fig:3samegene_a_2}, main products of about 1000 bp and a faint band of about 2000 bp is visible, while expected concatenations in all five reactions were $\alpha A \gamma A \beta$ (2101 bp) and $\alpha A \gamma A \gamma A \beta$ (3141 bp).

This result has been verified in several experimental conditions, such as different DNA polymerases
(\gene{Taq} and \gene{Pfu}), different annealing temperatures, and different template ratio concentrations (1:1, 1:2, 1:5, 1:10). Due to this phenomenon, a major limit has been experimentally outlined for XPCR working on fragments containing long identical sequences. Indeed, after these experiments, we may claim that over molecules containing the same fragment, the XPCR$_\gamma$ procedure realizes a string operation different than the expected overlap assembly, a sort of ``fusion operation'', defined as: $\alpha X \gamma, \gamma X \beta \; \rightarrow \;  \alpha X \beta$.
 
\begin{figure}[h!] 
	\centering
	\includegraphics[width=.69\columnwidth]{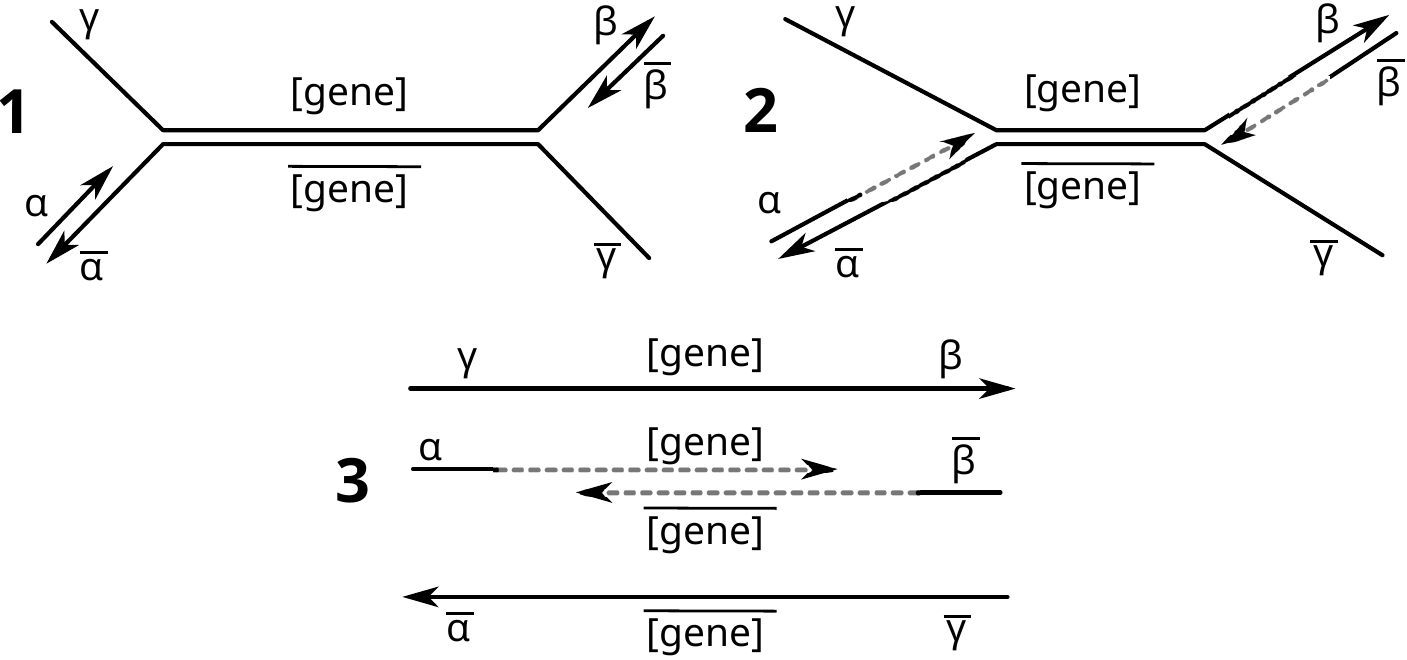}
	\caption{Amplification of new molecules $\alpha X \beta$, X one of the genes, by strand displacement.}
	\label{fig:alpha-gene-beta}
\end{figure}

We hypothesize that the above phenomenon is caused by strand displacement (template switching and/or
hydrolysis of competing strands)~\cite{Kanagawa:03}, a graphical explanation of which is reported in
Figure~\ref{fig:alpha-gene-beta}. 

Starting from $\alpha X \gamma$ and $\gamma X \beta$, in Figure~\ref{fig:alpha-gene-beta} (1) a heteroduplex is formed after denaturation, by annealing along the gene X resulting in two flap structures, partially complementary to the primers $\alpha$ and $\overline{\beta}$. In (2) primers anneal and DNA polymerase reaches the double strand of the heteroduplex. Here the enzyme cuts the flapping $\gamma$-containing sequence, due to its 5$^\prime$$\rightarrow$3$^\prime$ exonuclease activity~\cite{Ceska:98} (aided by the fork-like structure
	\cite{Holland:91}\cite{Pavlov:04} and by proximity with primers \cite{Lyamichev:93}), while the extension of
	both strands continues. They are elongated while nick translation removes bases from the opposite
	template strand, until (3) both polymerases meet around the middle region of the gene, where a template switch occurs and DNA
	replication completes the formation of the two chimeric molecules~\cite{Kanagawa:03}. After denaturation and subsequent annealing, a simple XPCR (super-)amplifies $\alpha X \beta$.

\section{Multiple XPCR based three genes concatenation}\label{sec:3}

\begin{figure}[ht] 
	\centering
	\includegraphics[width=.99\columnwidth]{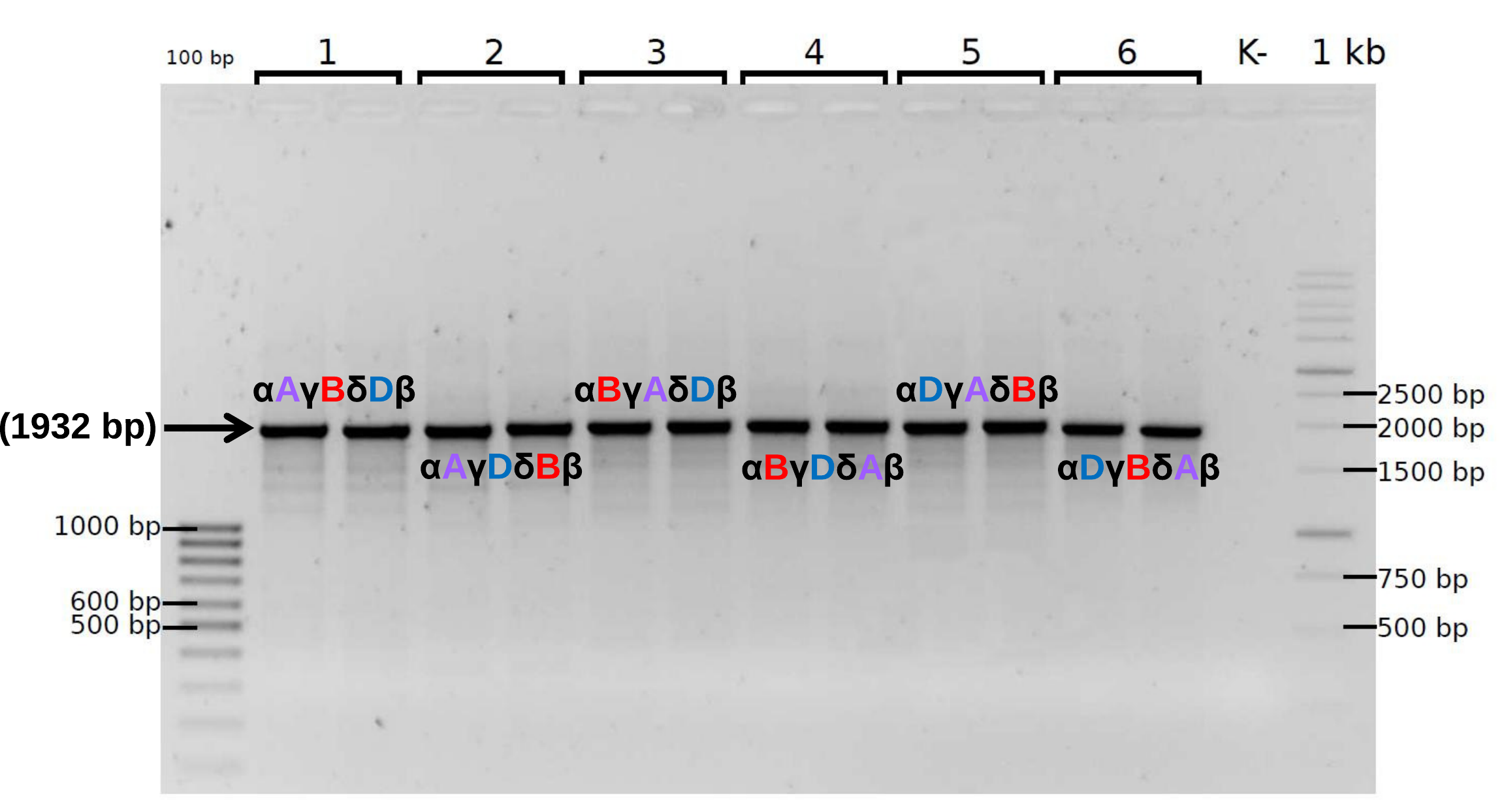}
	\caption{Successful multiple XPCR to concatenate three genes. Lanes 1, 2, 3, 4, 5, 6: templates $\alpha A \gamma$ + $\gamma B \delta$ +
	$\delta D \beta$, $\alpha A \gamma$ + $\gamma D \delta$ + $\delta B \beta$, $\alpha B \gamma$ +
	$\gamma A \delta$ + $\delta D \beta$, $\alpha B \gamma$ + $\gamma D \delta$ + $\delta A \beta$, $\alpha D
	\gamma$ + $\gamma A \delta$ + $\delta B \beta$, and $\alpha D \gamma$ + $\gamma B \delta$ + $\delta A \beta$,
	respectively. Lane K-: negative control without any template. For experimental details see Table~\ref{tab1}.}
	\label{fig:gamma2_2}
\end{figure}

Here we prove that multiple $XPCR$  works correctly to concatenate three genes of different length. Experimental results are reported in Figure~\ref{fig:gamma2_2}, where permutations of genes have been produced in one step, by PCR reactions with primers ($\alpha$, $\overline{\beta}$). This achievement represents the validation of multiple XPCR on natural long sequences, since the technique in the literature has been tested only for short and artificial sequences. 

\begin{figure}[h] 
	\centering
	\includegraphics[width=.89\columnwidth]{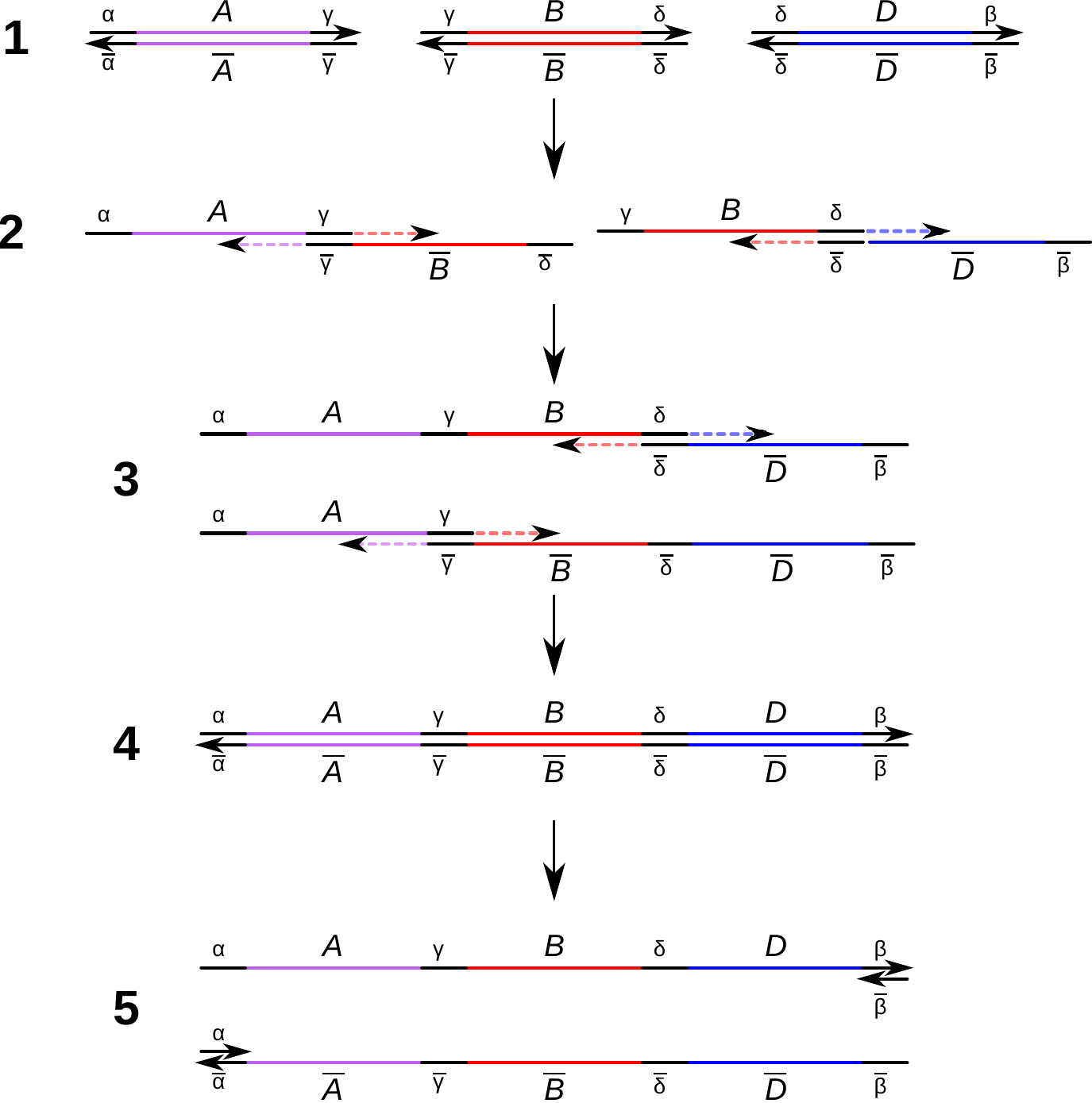}
	\caption{Three genes concatenation from $\alpha A \gamma$ + $\gamma B \delta$ + $\delta D \beta$ in one step. (1)
	Starting templates. (2-3) Annealing along common $\gamma$ and $\delta$, and extension by DNA polymerase, leading to concatenation of genes A, B, and D in (4) which is amplified in (5).}
	\label{fig:3genedelta_drw}
\end{figure}

According to the $n$-XPCR method~\cite{Franco:11}, if $n$ is set equal to 3 and a PCR with primers
($\alpha$, $\overline{\beta}$) is performed on a starting pool P containing the three different
sequences $\{\alpha \delta_1 \gamma_1, \gamma_1 \delta_2 \gamma_2, \gamma_2 \delta_3 \beta\}$, the product
$\alpha \delta_1 \gamma_1 \delta_2 \gamma_2 \delta_3 \gamma_3 \beta$ is amplified (as in Figure~\ref{fig:3genedelta_drw}). Successful results are reported in Figure~\ref{fig:gamma2_2}, where all lanes exhibit the expected products (molecules 1932 bp long), where we have instantiated $\delta_1, \delta_2, \delta_3$ with our genes, $\gamma_1 = \gamma$ and $\gamma_2 = \delta$. 

Beside, an attempt was performed to implement the special case of $\gamma = \delta$, in order to possibly simplify the (requred design and the) implementation of multiple XPCR. This attempt showed, as it is described in the following subsection, that the one-step library generation by XPCR cannot be further simplified, and it requires all different overlap primers (which is a limit for the scale-up of the procedure, for primer design constraints).

\subsection{Attempt with one only overlapping primer (case $\gamma = \delta$)} \label{sub3}

\begin{figure}[h] 
	\centering
	\includegraphics[width=.99\columnwidth]{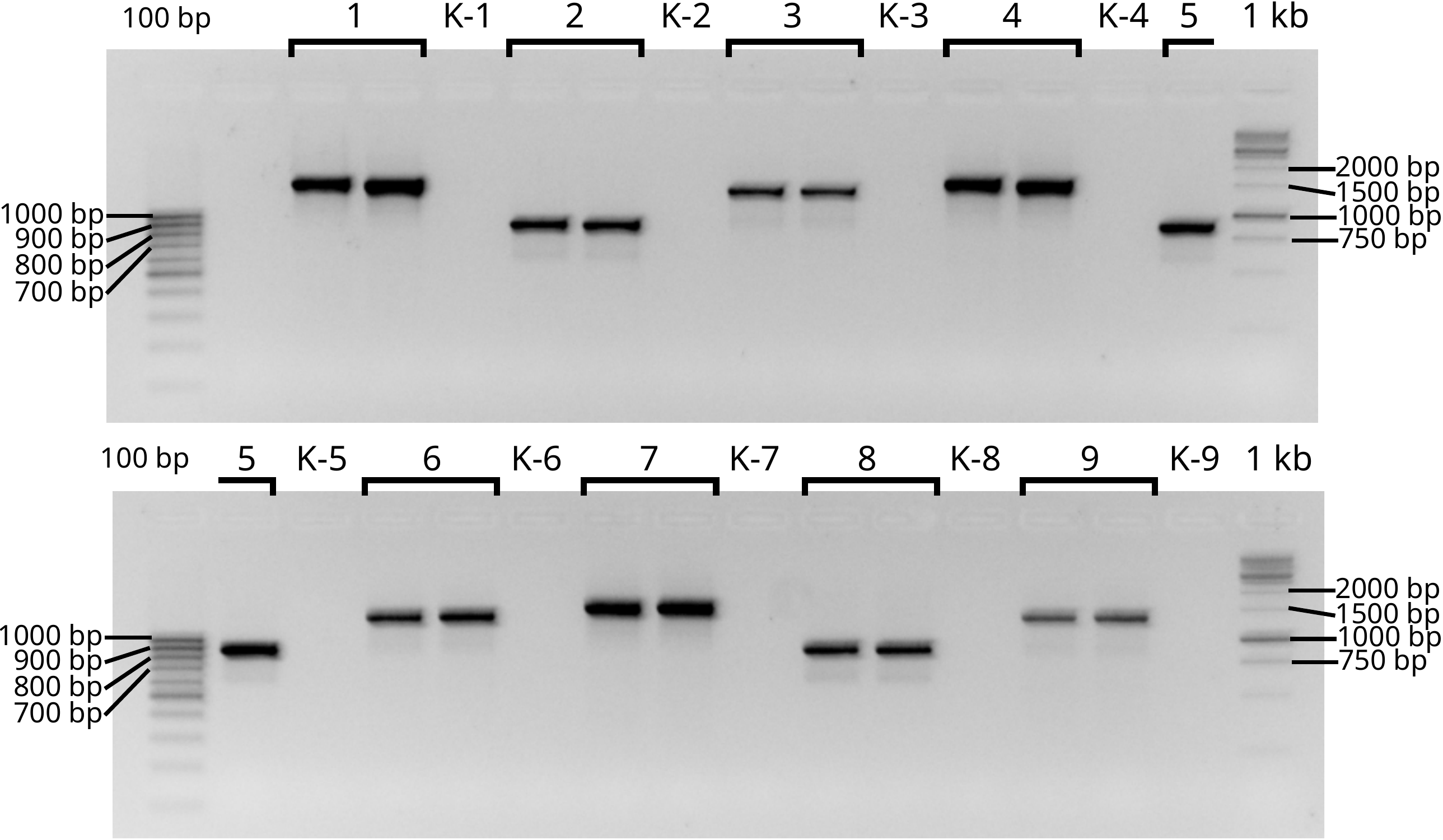}
	\caption{Multiple XPCR may not be further simplified. Lanes 1, 4, 7: template $\alpha A \gamma$ + $\gamma B \gamma$ + $\gamma D \beta$; concentration ratio of $\gamma B \gamma$ to the other templates is, respectively, 4:1, 8:1 and 16:1; exhibited products of about 1600 bp.
	Lanes 2, 5, 8: templates $\alpha B \gamma$ +
	$\gamma A \gamma$ + $\gamma D \beta$; concentration ratio of $\gamma A \gamma$ to the other templates is,
	respectively, 4:1, 8:1 and 16:1; exhibited products of about 900 bp.
	Lanes 3, 6, 9: templates $\alpha B \gamma$ + $\gamma D \gamma$ + $\gamma A \beta$; concentration ratio of $\gamma D \gamma$ to the other templates is,
	respectively, 4:1, 8:1 and 16:1; exhibited products of about 1400 bp.
	Lanes K-1, K-2, K-3, K-4, K-5, K-6, K-7, K-8, K-9: negative controls without any template, for reactions in lanes
	1, 2, 3, 4, 5, 6, 7, 8, 9 respectively. For experimental details see Table~\ref{tab2}.}
	\label{fig:3gene}
\end{figure}

For the sake of economy, it has been attempted to concatenate our three different genes by multiple XPCR, with equal overlapping primers $\gamma$ and $\delta$. PCR reactions with primers ($\alpha$,
	$\overline{\beta}$) over  templates $\alpha X \gamma$ + $\gamma Y \gamma$ + $\gamma Z \beta$, with X, Y, Z three different genes, may in principle produce two different amplicons $\alpha X \gamma Z \beta$ (by basic XPCR) and $\alpha X \gamma Y \gamma Z \beta$ (by multiple XPCR), while in all the experiments only the first one has been (consistently) produced. Results are shown in Figure~\ref{fig:3gene}, where only shortest products (that is, the concatenation of two genes) were amplified, and no longer products (of expected length 1932 bp) are visible, despite the different concentration ratios (4:1, 8:1 and 16:1) in each sample between $\gamma X \gamma$ templates ($X$ one of the genes)  and the other templates $\alpha X \gamma$ and $\gamma X \beta$. This is probably caused by the earlier formation of the shortest amplification seed.


As a conclusion, we have found out that $n$-XPCR does extend basic XPCR with multiple concatenation (of $n$ fragments, even of different lengths) if and only if the $n-1$ overlapping primers are all different each other, and this claim has been experimentally tested for $n=3$.

\section{Conclusions}

An experimental analysis of both basic and multiple XPCR protocols, previously introduced~\cite{Franco:05b,Franco:05,Franco:06,Manca:08,Franco:11}, has been carried out in this work, where limits and potentialities of the techniques to concatenate several genes have been outlined. As a biological setting, we have worked with the three genes \gene{dbtAa} (1019 bp, ferredoxyn reductase), \gene{dbtAb} (311 bp, ferredoxyn), and \gene{dbtAd} (518 bp, $\beta$ dioxygenase subunit), extracted from the bacterial strain \specie{Burkholderia fungorum} DBT1 ~\cite{DiGregorio:04},  widely studied for its notable PAH degradation capabilities. All experiments have been performed in double sampling, with negative controls, under different experimental conditions (including temperature, concentration, gene and length variations), and repeated with both {\it Taq} and {\it Pfu} polymerases. Experimental details are reported in the following appendix section of Materials and Methods.

XPCR technique worked as expected when concatenating two genetic sequences of different lengths, independently from their length. Moreover, by applying multiple XPCR, we were also able to recombine all possible permutations of three genes, in one step and with high yields. Aside it was interestingly demonstrated that by applying XPCR or multiple XPCR methods, it is not possible to concatenate multiple occurrences of the same gene, in an appreciable quantity. This result is likely due to a strand displacement phenomenon which happens when performing a PCR with primers ($\alpha$, $\overline{\beta}$) on sequences $\alpha X \gamma$ and $\gamma X \beta$, which results in the only amplicon $\alpha X \beta$. This is a special case where overlap assembly cannot be implemented by XPCR, which instead realizes a different operation (still to be investigated by a formal language viewpoint): a sort of molecular fusion of the two input. Such an experimental evidence solicits a refined extension of combinatorial models for DNA polymerases and PCR based procedures from the literature~\cite{Manca:08,book}.

Finally, experimental results reported in subsection~\ref{sub3} limit the scale-up of multiple XPCR, due to primer design constrains~\cite{brodin:13}. However this technique is still one of the most efficient (logarithmic time with the number of variables to recombine~\cite{Franco:05}) and reliable way to generate large DNA libraries~\cite{libraries}.

\appendix

\section{Materials and methods}

{\bf Reagents}: Sharpmass\textsuperscript{\texttrademark} 1 (0.25 to 1 kb) and Sharpmass\textsuperscript{\texttrademark} 100 (100 to 1000 bp) were both purchased from EuroClone S.p.A (Italy). PCR buffer, MgCl2, dNTP, GoTaq$\circledR$ DNA Polymerase and Pfu DNA Polymerase were furnished by Promega (Milan, Italy). All the synthetic DNA oligonucleotides and all the primers were from Sigma-Aldrich.\\

\noindent{\bf Bacterial genes}: Gene fragments were extracted from {\it Burkholderia fungorum} DBT1~\cite{DiGregorio:04}, an environmental bacterial isolate with remarkable PAH (polycyclic aromatic hydrocarbon) degrading capabilities. In particular three catabolic genes coding for three subunits of the initial dioxygenase were used. They were \gene{dbtAa}(1019 bp) encoding for ferredoxyn reductase, \gene{dbtAb} (311 bp) encoding for ferredoxyn, and \gene{dbtAd} (518 bp) encoding for dioxygenase subunit $\beta$. Gene sequences are available in GenBank with accession number AF380367 for gene \gene{dbtAd} and AF404408 for genes \gene{dbtAa} and \gene{dbtAb} (in this work \gene{dbtAa} is denoted by $A$, \gene{dbtAb} by $B$, and \gene{dbtAd} by $D$). \\

\noindent{\bf Primers design and PCR conditions}: All the primers~\footnote{\scriptsize $\alpha$ =
5$^\prime$-\seqsplit{TTCTACAAGGAGGATATTACC}-3$^\prime$, $\overline{\beta}$ =
5$^\prime$-\seqsplit{TATGGAGATGTACCTGATATC}-3$^\prime$, $\gamma$ =
5$^\prime$-\seqsplit{ATATTGGAGGAGGTATACAAC}-3$^\prime$, $\overline{\gamma}$ =
5$^\prime$-\seqsplit{GTTGTATACCTCCTCCAATAT}-3$^\prime$, $\delta$ =
5$^\prime$-\seqsplit{GAATTACAGGAGGATGTTTGG}-3$^\prime$, $\overline{\delta}$ =
5$^\prime$-\seqsplit{CCAAACATCCTCCTGTAATTC}-3$^\prime$.} and oligos used in this study were designed and checked with the aid of MATLAB$\circledR$ (The MathWorks, Inc.) and its Bioinformatics toolbox\textsuperscript{\texttrademark} .

All PCR reactions, except otherwise stated, were carried out in 25 \SI{}{\micro\liter} of total volume containing 0.8 $\mu$M of each primer ($\alpha$ and $\beta$), 0.4 mM of dNTPs, 2.5 U of $Pfu$ DNA polymerase (Promega, Milan, Italy) and 2.5 \SI{}{\micro\liter} of 10x PCR buffer. Concentration of template was 20 ng per reaction.

In the case of XPCR-based two genes concatenation experiments, amplifications were performed over the following couple of templates: $\alpha A \gamma$/$\gamma B \beta$; $\alpha B \gamma$/$\gamma D \beta$; $\alpha D \gamma$/$\gamma B \beta$; $\alpha D \gamma$/$\gamma A \beta$. PCR thermocycler conditions were as follows: 94$\si{\celsius}$ for 2 min, then 30 cycles of 94$\si{\celsius}$ for 1 min, 51$\si{\celsius}$ for 45 sec and 72$\si{\celsius}$ for 4 min, with a final extension step at 72$\si{\celsius}$ for 5 min.

Attempt to concatenate two and three copies of a gene. In the case of concatenation of two copies of the same gene, amplification was performed over $\alpha D \gamma$/$\gamma D \beta$ templates. PCR thermocycler conditions were: 94$\si{\celsius}$ for 2 min, then 30 cycles of 94$\si{\celsius}$ for 1 min, 49$\si{\celsius}$ for 1 min and 72$\si{\celsius}$ for 2 min, with a final extension step at 72$\si{\celsius}$ for 5 min. For the concatenation of three copies of the same gene, amplification was performed over $\alpha A \gamma$/$\gamma A \gamma$/$\gamma A \beta$ genes with the following template concentrations: 10 ng/reaction for $\alpha A \gamma$ and $\gamma A \beta$, and 100 ng/reaction for $\gamma A \gamma$. Experiment was carried out at different annealing temperature ($t_1$=47.9; $t_2$=48.7; $t_3$=50.7; $t_4$=51.8; $t_5$=53.8). Thermocycler conditions were as follows: 94$\si{\celsius}$ for 2 min, then 30 cycles of 94$\si{\celsius}$ for 1 min, 1 min of incubation at different annealing temperatures and 72$\si{\celsius}$ for 4 min, with a final extension step at 72$\si{\celsius}$ for 5 min.

Successful multiple XPCR to concatenate three genes. In this case, PCR reactions were carried out in 25 \SI{}{\micro\liter} of total volume containing 0.8 $\mu$M of each primer ($\alpha$ and $\beta$), 0.4 mM of dNTPs, 1.25 U of GoTaq$\circledR$ DNA Polymerase and 2.5 \SI{}{\micro\liter}  of 5x PCR buffer, with templates reported in Table~\ref{tab1}. 
PCR thermocycler conditions were as follows: 94$\si{\celsius}$ for 2 min, then 30 cycles of 94$\si{\celsius}$ for 45 sec, 55$\si{\celsius}$ for 30 sec and 72$\si{\celsius}$ for 2 min and 30 sec, with a final extension step at 72$\si{\celsius}$ for 2 min and 30 sec.

Multiple XPCR may not be further simplified. PCR reactions were carried out with GoTaq$\circledR$ DNA Polymerase (with the same reaction mixture of the previous experiment), with templates reported in Table~\ref{tab2}. 
The PCR thermocycler conditions were as follows: 94$\si{\celsius}$ for 2 min, then 30 cycles of 94$\si{\celsius}$ for 45 sec, 51$\si{\celsius}$ for 30 sec and 72$\si{\celsius}$ for 4 min, with a final extension step at 72$\si{\celsius}$ for 5 min. \\

{\bf Gel Electrophoresis}: Agarose gels were prepared at 1 or 2 \% (w/vol) on the basis of expected dimensions of PCR products. Electrophoretic runs were carried out in TAE (1X) at 10 volt/cm$^2$. Sharpmass\textsuperscript{\texttrademark}  1 (0.25 to 1 kb) and Sharpmass\textsuperscript{\texttrademark}  100 (100 to 1000 bp) were used as DNA mass ladders. The presence of bands was detected by a digital gel scanner. The DNA bands (final PCR products) of interest were excised from the gel and further purified through QIAEX$\circledR$ II Gel Extraction Kit (QIAGEN, US) following the manufacturers instructions. Finally, eluted DNA fragments were sequenced on both strands.


\begin{table}[ht] 
\begin{center}
\scalebox{0.95}{
\begin{tabulary}{\linewidth}{*{7}{C}}
	\toprule
	\multirow{2}{*}{\bfseries Component} & \multicolumn{6}{c}{\bfseries Quantity (\SI{}{\micro\liter})} \\
	\cmidrule(lr){2-7}
	& \bfseries \textit{ABD} & \bfseries \textit{ADB} & \bfseries \textit{BAD} & \bfseries
	\textit{BDA} & \bfseries \textit{DAB} & \bfseries \textit{DBA}
	\\
	\cmidrule(lr){1-1} \cmidrule(lr){2-7}
	Template $\alpha A \gamma$ & 2 & 2 & - & - & - & - \\
	Template $\alpha B \gamma$ & - & - & 2 & 2 & - & - \\
	Template $\alpha D \gamma$ & - & - & - & - & 2 & 2 \\
	Template $\gamma A \delta$ & - & - & 2 & - & 2 & - \\
	Template $\gamma B \delta$ & 2 & - & - & - & - & 2 \\
	Template $\gamma D \delta$ & - & 2 & - & 2 & - & - \\
	Template $\delta A \beta$ & - & - & - & 2 & - & 2 \\
	Template $\delta B \beta$ & - & 2 & - & - & 2 & - \\
	Template $\delta D \beta$ & 2 & - & 2 & - & - & - \\
	\bottomrule
\end{tabulary}}
\end{center}
	\caption{PCR components for three genes concatenation, by multiple XPCR with $\gamma \not = \delta$. }
	\label{tab1}
\end{table}
\vspace{-0.95cm}
\begin{table}[htb] 
\begin{center}
\scalebox{0.97}{
\begin{tabulary}{\linewidth}{*{10}{C}}
	\toprule
	\multirow{2}{*}{\bfseries Component} & \multicolumn{9}{c}{\bfseries Quantity (\SI{}{\micro\liter})} \\
	\cmidrule(lr){2-10}
	& \bfseries 1.1 & \bfseries 1.2 & \bfseries 1.3 & \bfseries
	2.1 & \bfseries 2.2 & \bfseries 2.3 & \bfseries 3.1 & \bfseries 3.2 & \bfseries 3.3 \\
	\cmidrule(lr){1-1} \cmidrule(lr){2-4} \cmidrule(lr){5-7} \cmidrule(lr){8-10}
	Template $\alpha A \gamma$ & 0.5 & - & - & 0.5 & - & - & 0.5 & - & - \\
	Template $\alpha B \gamma$ & - & 0.5 & 0.5 & - & 0.5 & 0.5 & - & 0.5 & 0.5 \\
	Template $\gamma A \gamma$ & - & 2 & - & - & 4 & - & - & 8 & - \\
	Template $\gamma B \gamma$ & 2 & - & - & 4 & - & - & 8 & - & - \\
	Template $\gamma D \gamma$ & - & - & 2 & - & - & 4 & - & - & 8 \\
	Template $\gamma A \beta$ & - & - & 0.5 & - & - & 0.5 & - & - & 0.5 \\
	Template $\gamma D \beta$ & 0.5 & 0.5 & - & 0.5 & 0.5 & - & 0.5 & 0.5 & - \\
	\bottomrule
\end{tabulary}}
\end{center}
	\caption{PCR components for three genes concatenation,  by multiple XPCR with $\gamma = \delta$.}
	\label{tab2}
\end{table}

\end{document}